\documentclass[reprint,superscriptaddress,amssymb,amsmath,aps,showpacs,10pt,floatfix,prl,longbibliography]{revtex4-2}
\def\FeSn{Fe$_3$Sn$_2$}
\def\cm{cm$^{-1}$}
\usepackage{graphicx}%
\usepackage{color}
\usepackage{epstopdf}
\usepackage{amssymb}
\usepackage{amsmath}
\usepackage{amsfonts}
\usepackage{sidecap}
\sidecaptionvpos{figure}{t}

\usepackage{color}
\usepackage[colorlinks,bookmarks=false,citecolor=darkblue,linkcolor=red,urlcolor=blue]{hyperref} 
\definecolor{darkred}{rgb}{0.7,0.0,0.0}

\definecolor{darkblue}{rgb}{0,0.02,0.45}

\definecolor{darkgreen}{rgb}{0.02,0.45,0.0}

\definecolor{violet}{rgb}{0.8,0.2,0.6}

\begin{document}
\title{Coherent phonon and unconventional carriers in the magnetic kagome metal \FeSn}

\author{M. V. Gonçalves-Faria}
\email{m.goncalves-faria@hzdr.de}
\affiliation{Institute of Ion Beam Physics and Materials Research, Helmholtz-Zentrum Dresden-Rossendorf, 01328 Dresden, Germany}
\affiliation{Technische Universität Dresden, 01062 Dresden, Germany}

\author{A. Pashkin}
\affiliation{Institute of Ion Beam Physics and Materials Research, Helmholtz-Zentrum Dresden-Rossendorf, 01328 Dresden, Germany}

\author{Q. Wang}
\affiliation{Department of Physics and Beijing Key Laboratory of Opto-electronic Functional Materials $\&$ Micro-nano Devices, Renmin University of China, Beijing 100872, China}

\author{H. C. Lei}
\affiliation{Department of Physics and Beijing Key Laboratory of Opto-electronic Functional Materials $\&$ Micro-nano Devices, Renmin University of China, Beijing 100872, China}

\author{S. Winnerl}
\affiliation{Institute of Ion Beam Physics and Materials Research, Helmholtz-Zentrum Dresden-Rossendorf, 01328 Dresden, Germany}

\author{A. A. Tsirlin}
\affiliation{Felix Bloch Institute for Solid-State Physics, Leipzig University, 04103 Leipzig, Germany}

\author{M. Helm}
\affiliation{Institute of Ion Beam Physics and Materials Research, Helmholtz-Zentrum Dresden-Rossendorf, 01328 Dresden, Germany}
\affiliation{Technische Universität Dresden, 01062 Dresden, Germany}

\author{E. Uykur}
\email{e.uykur@hzdr.de}
\affiliation{Institute of Ion Beam Physics and Materials Research, Helmholtz-Zentrum Dresden-Rossendorf, 01328 Dresden, Germany}

\date{\today}

\begin{abstract}
Temperature- and fluence-dependent carrier dynamics of the magnetic Kagome metal \FeSn\ were studied using the ultrafast optical pump-probe technique. Two carrier relaxation processes ($\tau_1$ and $\tau_2$) and a laser induced coherent optical phonon were observed. By using the two-temperature model for metals, we ascribe the shorter relaxation $\tau_1$ ($\sim$1~ps) to hot electrons transferring their energy to the crystal lattice via electron-phonon scattering. $\tau_2$ ($\sim$25~ps), on the other hand, cannot be explained as a conventional process and is attributed to the unconventional (localized) carriers in the material. The observed coherent oscillation is assigned to be a totally symmetric $A_{1g}$ optical phonon dominated by Sn displacements out of the Kagome planes, and possesses a prominently large amplitude, on the order of 10$^{-3}$, comparable to the maximum of the reflectivity change ($\Delta$R/R). This amplitude is equivalent to charge-density-wave (CDW) systems, although no signs of such an instability were hitherto reported in \FeSn. Our results set an unexpected connection between \FeSn\ and kagome metals with CDW instabilities, and suggest a unique interplay between phonon and electron dynamics in this compound.
\end{abstract}

\maketitle

\textcolor{blue}{\textit{Introduction.}} During the last years magnetic Kagome metals have emerged as an interesting new class of materials due to their unusual properties. In terms of electronic structure, a simple tight-binding model constructed on the Kagome lattice is well known to result in dispersionless flat bands and Dirac cones~\cite{DiracFermionsAndFlatBands}. Thus, strongly correlated and localized electrons together with topological states are expected for these materials. Combining such features with magnetism makes Kagome metals suitable to host different types of exotic phenomena, and in this regard the family of Kagome FeSn-binary compounds (FeSn, Fe$_3$Sn, \FeSn) presents several promising candidates. For \FeSn\ both the linearly dispersing bands and the flat bands were previously observed experimentally \cite{DiracFermions,Lin2018Aug}, and recently it attracted great attention after discoveries of massive Dirac fermions \cite{DiracFermions}, large anomalous Hall effect \cite{GiantHallEffect}, skyrmion bubbles at room temperature \cite{SkyrmionsRoomT} and tunable spin textures using an external magnetic field \cite{Altthaler2021Dec}.

\begin{figure}
	\centering
	    \includegraphics[width=\columnwidth]{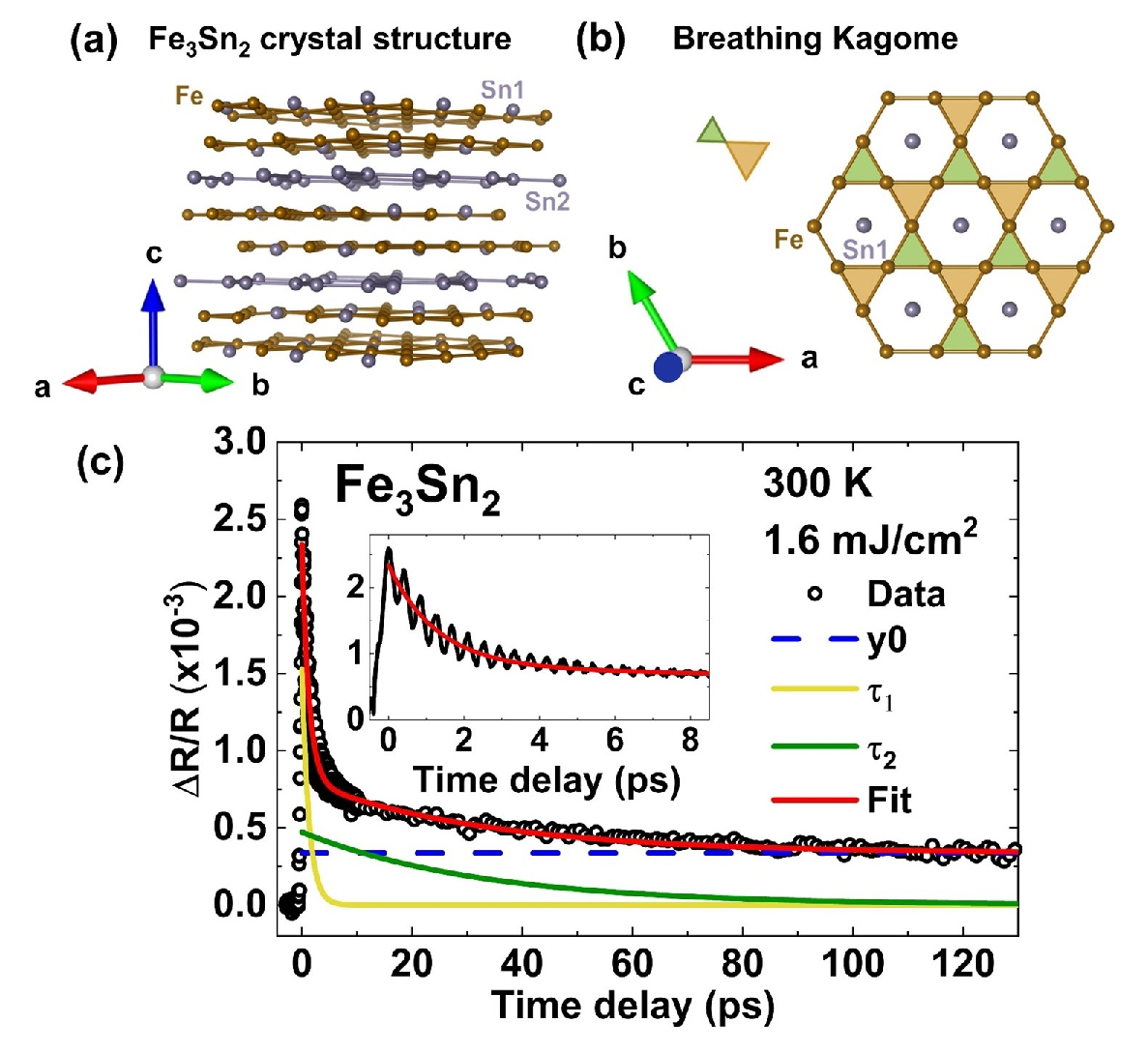}
	\caption{(a) Crystal structure of \FeSn, with Fe$_3$Sn1 kagome bilayers and honeycomb Sn2 layers. (b) Fe-kagome network centered by the Sn1 atoms and the breathing kagome bonds demonstrated with triangles of different color. (c) Optical pump-probe spectrum of \FeSn\ at $300$~K with the fluence of $1.6$~mJ/cm$^2$. Black dots are experimental data, and red solid line is the exponential fit according to Eq.~\ref{E1}. Blue dashed line is the offset constant $y0$, which is an approximation for a long cooling down process that is out of our time delay. Yellow and green solid lines represent $\tau_1$ and $\tau_2$ exponential decay processes, respectively. The inset shows a zoom to the first 8 picoseconds, where pronounced oscillations were observed in all measured temperatures. These are coherent phonon oscillations with resonance frequency corresponding to one of the $A_{1g}$ totally symmetric phonon modes of \FeSn.} 
	\label{F1}
\end{figure}

\FeSn\ is a layered rhombohedral material belonging to the $R\bar{3}m$ space group, with hexagonal lattice parameters $a = b = 5.3$ \r A and $c = 19.8$ \r A. Its crystalline structure is composed of Fe$_3$Sn kagome bilayers, where the Fe kagome network is stabilized with Sn1 atoms and sandwiched between honeycomb Sn2 layers [Fig.~\ref{F1}(a,b)]~\cite{LatticeParameters}. Previous studies reported a high ferromagnetic ordering temperature, T$_C \sim 640$~K~\cite{LatticeParameters,Mosbauer_TC,Mosbauer_TC2}, and identified a temperature-driven spin reorientation in \FeSn. The spin reorientation, where the spins are realigned from the out-of-plane direction towards in-plane \cite{Mosbauer_TC}, occurs in a broad temperature range, between $150$~K and $70$~K, and its signatures were observed with several different experimental techniques, such as Mössbauer \cite{Mosbauer_SpinR,Mosbauer_TC2}, neutron diffraction \cite{LatticeParameters}, electronic transport \cite{Wang2016Aug} and infrared spectroscopy \cite{Biswas2020Aug}. Furthermore, due to the bilayer nature of the compound and the presence of a breathing Kagome distortion on the Fe-Kagome layer [Fig.~\ref{F1}(b)], a slightly modified band structure is expected \cite{DiracFermions}.

It has been shown that the observed properties of \FeSn\ are closely related to peculiarities of its lattice and magnetic structure \cite{Tanaka2020Apr,DiracFermions}. The fingerprints of the non-trivial carrier dynamics have been identified in optical studies \cite{Biswas2020Aug}, whereas the  interplay of the topological orders with magnetism and strongly correlated electrons is yet to be clarified. The tunability of different contributions is highly desirable, also for possible future applications of \FeSn.

Here, we present an ultrafast optical pump-probe spectroscopy investigation on \FeSn. This method has been extensively used to study the dynamics of non equilibrium charge carriers and phonon dynamics in solids \cite{Ultrafast1,Ultrafast3,Ultrafast4,Misochko2002Jul,CoherentPhononsAntimony}, and it is well suited to study metallic systems  \cite{Ultrafast_metals1,Ultrafast_metals2,Ultrafast_metals3,Ultrafast_metals4}, where different contributions can be identified. So far, the ultrafast carrier dynamics for Kagome metals have not yet been widely explored, with only a few reports on nonmagnetic CsV$_3$Sb$_5$ \cite{Wang2021Oct, Ratcliff2021Nov}, where the ultrafast response of the unusual charge-density-wave (CDW) state has been probed. In this letter, we report the temperature- and fluence-dependent transient reflectivity measurements of \FeSn\ using optical pump-probe. Our results reveal an unusually large amplitude of coherent phonon oscillations in \FeSn, with intriguing similarities to the CDW case, even though no CDW has been reported in \FeSn\ as a ferromagnetic Kagome metal. Thus, giving new insights into the electron-phonon coupling as the possible mechanism related to the unconventional carriers in kagome metals.

\textcolor{blue}{\textit{Experimental.}} A temperature- and fluence- dependent optical pump-probe spectroscopy study has been performed on as-grown \FeSn\ single crystals \cite{Wang2016Aug} in the reflection geometry. For pump and probe we used $\sim 60$~fs long laser pulses, centered at $800$~nm and generated by a Ti:sapphire laser amplifier with $250$~kHz repetition rate. Further experimental details can be found in the supplementary material. 

In Fig.~\ref{F1}(c), we summarize the general behavior of the observed relaxation processes. Here, reflectivity change ($\Delta R/R$) is given as a function of the pump-probe time delay at $300$~K with pump fluence of $1.6$~mJ/cm$^2$. The best fit for the spectrum was achieved with the following equation:
\begin{equation}
    \centering
    \Delta R/R = y0 + c_1\exp{(-t/\tau_1)} + c_2\exp{(-t/\tau_2)},
    \label{E1}
\end{equation}
where $c_1$ and $c_2$ are constants, $y0$ is an offset parameter and $\tau_1$ and $\tau_2$ are the relaxation times. The time scales of the relaxations were: $\tau_1$ in the order of $\sim 1$~ps (yellow solid line), $\tau_2$ in the order of a few tens of picoseconds (green solid line), and finally a much longer relaxation that had to be approximated using the offset constant term $y0$ (blue dashed line). Another interesting feature is that around the first $\sim  8$~ps after pump-probe temporal overlap, the decaying signal is modulated by pronounced oscillations, as seen in the inset of Fig.~\ref{F1}(c). This is a coherent optical phonon induced by the ultrashort pump pulse and will be discussed in more detail later on. 

\begin{figure*}
	\centering
	    \includegraphics[width=2\columnwidth]{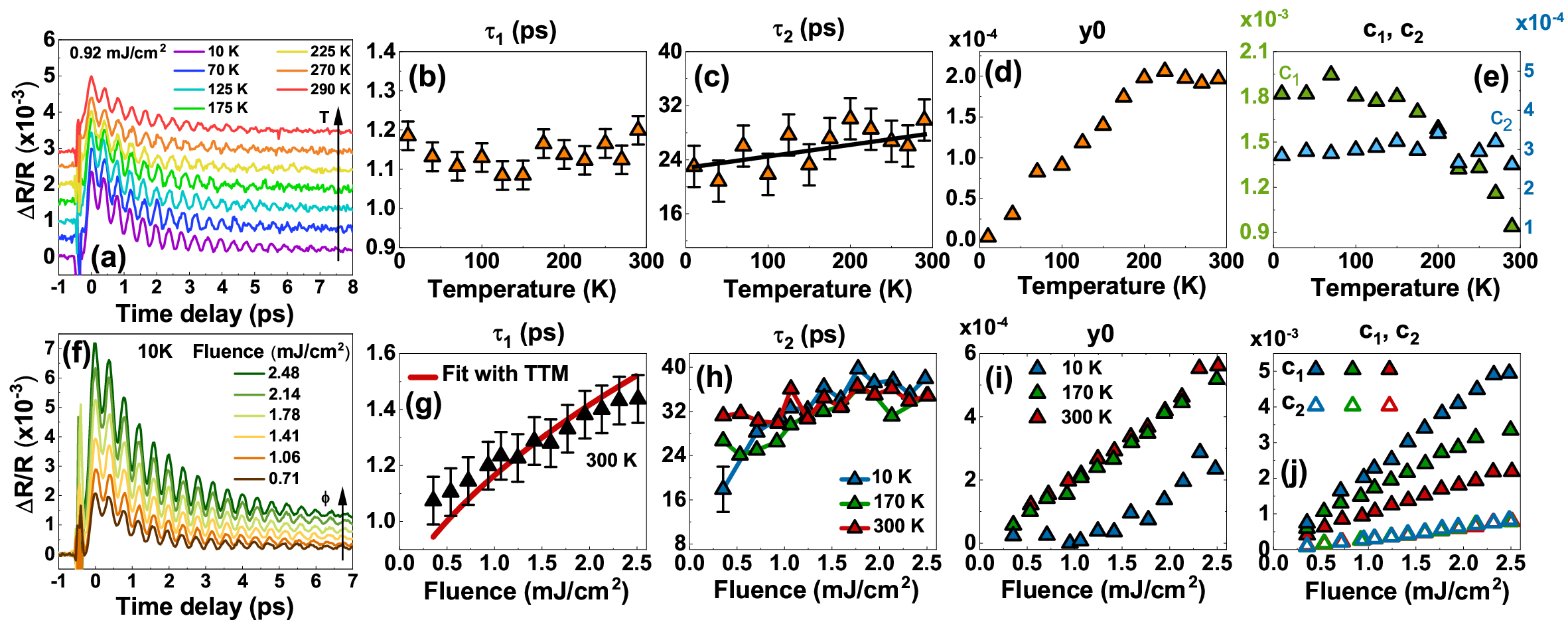}
	    \caption{(a) Temperature-dependent optical pump-probe spectra of \FeSn\ with $0.92~$mJ/cm$^2$ of pump fluence. The spectra at different temperatures are separated by a $0.5 \times 10^{-3}$ offset. (b), (c), (d) and (e) present the Eq.~\ref{E1} parameters $\tau_1$, $\tau_2$, $y0$, $c_1$ and $c_2$, respectively. Fluence-dependent experimental data at $10$~K are shown in (f). Panels (g-j) are $\tau_1$, $\tau_2$, $y0$, $c_1$ and $c_2$ as a function of pump fluence. Red solid line in (g) is the fit for $\tau_1$ using the two-temperature model at 300K. Fluence dependence results for $\tau_1$ at all temperatures are quite similar, so for better visualization the data and the fits at $170$ and $10$ K were added to the supplementary material.} 
	\label{F2}
\end{figure*}

\textcolor{blue}{\textit{Relaxations.}} The temperature dependence of the transient reflectivity, the obtained relaxation times and the offset constant $y0$ are given in Fig.~\ref{F2}(a-d), whereas Fig.~\ref{F2}(e) depicts $c1$ and $c2$, the constants representing the amplitude of the $\tau_1$ and $\tau_2$ according to Eq.~\ref{E1}, respectively. Fig.~\ref{F2}(f-j) demonstrate the fluence dependence of the same parameters. We limited the time delay to 8~ps, longer time delays can be found in the supplementary material. 

Due to the metallic nature of \FeSn\ \cite{Wang2016Aug},  $\tau_1$ and $y0$ can be explained using the phenomenological two temperature model (TTM) for metals \cite{relaxation_copper,Ultrafast_metals1,Ultrafast_metals2}, where $\tau_1$ is the relaxation of the hot electrons, whereas $y_0$ reflects the dissipation of the residual lattice heating. As given in Fig.~\ref{F2}(b), $\tau_1$ is temperature independent, lying around $1.1$~ps. $y0$, on the other hand, increases with increasing temperature up to around $175$~K and then it saturates for higher temperatures, indicating that cooling down the sample removes the excess heat and brings the system to equilibrium faster [Fig.~\ref{F2}(d)]. The fluence dependencies of $\tau_1$ and $y0$ also corroborate this explanation as seen in Fig.~\ref{F2}(g) and (i), respectively. By simply taking into account the electron/lattice temperature and the electron-phonon coupling, the increase of $\tau_1$ with fluence can be nicely reproduced by the TTM model [red solid line in Fig.~\ref{F2}(g)]. A similar change has also been observed for 10~K and 170~K (see supplementary material for details of the TTM and the analysis for 10~K and 170~K). 

Coming to the $\tau_2$ as represented by the green solid line in Fig.~\ref{F1}, the dynamics behind this process indicates a departure from a simple Drude metal. Considering that the spectra are dominated by the coherent phonon oscillations and the excess heat of the system generates a background, $\tau_2$ is more reliably extracted at low temperatures, where $y0$ vanishes. The $\tau_2$ value is weakly temperature dependent [Fig.~\ref{F2}(c)] changing from 30~ps to $\sim$~25~ps with decreasing temperature. At high temperatures, we did not observe any fluence dependence [Fig.~\ref{F2}(h)]. With decreasing temperature at lower fluences, a small decrease is present and it goes into the saturation limit at higher pump fluences. 

We ascribe $\tau_2$ to the unconventional carriers that are expected in kagome metals. Previous optical studies~\citep{Biswas2020Aug} suggest that \FeSn\ is not a simple metal. Its optical conductivity shows two distinct intraband contributions. A sharp Drude contribution is accompanied by a second peak due to localized carriers (localization peak), which is the common situation on both magnetic and nonmagnetic kagome metals \cite{IR_study1,Biswas2020Aug,IR_study2,IR_study3,IR_study4}.  Here pumping leads to the delocalization of these unconventional carriers and we believe to observe the time scale of the localization process. The amplitude of this process should be proportional to the spectral weight of the localization peak observed in the broadband IR spectroscopy measurements~\cite{Biswas2020Aug}. Indeed a direct comparison reveals a temperature-independent behavior for both the spectral weight of the localization peak [Fig.S5(e) of Ref.~\cite{Biswas2020Aug}] and the amplitude of $\tau_2$ [$c_2$ in Fig.~\ref{F2}(e)].

The temperature-driven dynamics show a different evolution of $c_1$ and $c_2$ as given in Fig.~\ref{F2}(e). While with decreasing temperature, $c_2$ does not change, $c_1$ shows a slight increase and saturates below the spin-reorientation temperature. Here the change of the carrier density is probably not related with the change in the Fermi level with temperature, but rather with gapping of the certain parts of the Fermi surface upon the reorientation of the spins. With increasing fluence, on the other hand [Fig.~\ref{F2}(j)], a linear increase is observed for both $c_1$ and $c_2$, which is consistent with the increase of photo-excited carriers at higher fluences.

\textcolor{blue}{\textit{Phonon mode.}} Now let us turn to the coherent optical phonon identified in the spectra. These laser-induced oscillations are generated by the lattice atoms vibrating in phase to each other, and measured as a periodic modulation of the optical properties \cite{Coherent_Phonons_Theory,CoherentPhononsAntimony,CoherentPhononsBismuth}. In supplementary material the details regarding data analysis for the resonance frequency and amplitude of the mode can be found. Such coherent phonon oscillations are reported for different systems in the literature \cite{Optical_phonons1,Ultrafast_TI,softening_tellurium,CoherentPhononsAntimony,CoherentPhononsBismuth,phonon_bi2te3}. However, the extraordinary strength of these oscillations in the current measurements is an interesting finding. Such strong oscillations are usually observed in systems with periodic lattice distortions, charge density waves, and other types of collective order \cite{Tomeljak2009Feb,Schafer2010Aug,Kim2006Dec,coherent_phonon_cdw2,coherent_phonon_cdw3}, whereas \FeSn\ does not possess any of those. On the other hand, the correlated nature of the kagome metals has been identified by different means, including the observation of the aforementioned localization peak in the optical spectra. Here, the intraband carriers are damped by the back-scattering from the collective modes, which in principle can have any bosonic excitation as origin. Our observation of this unusual phonon coupling makes phonons a plausible candidate for this collective mode.

\begin{figure}
    \centering
        \includegraphics[width=1\columnwidth]{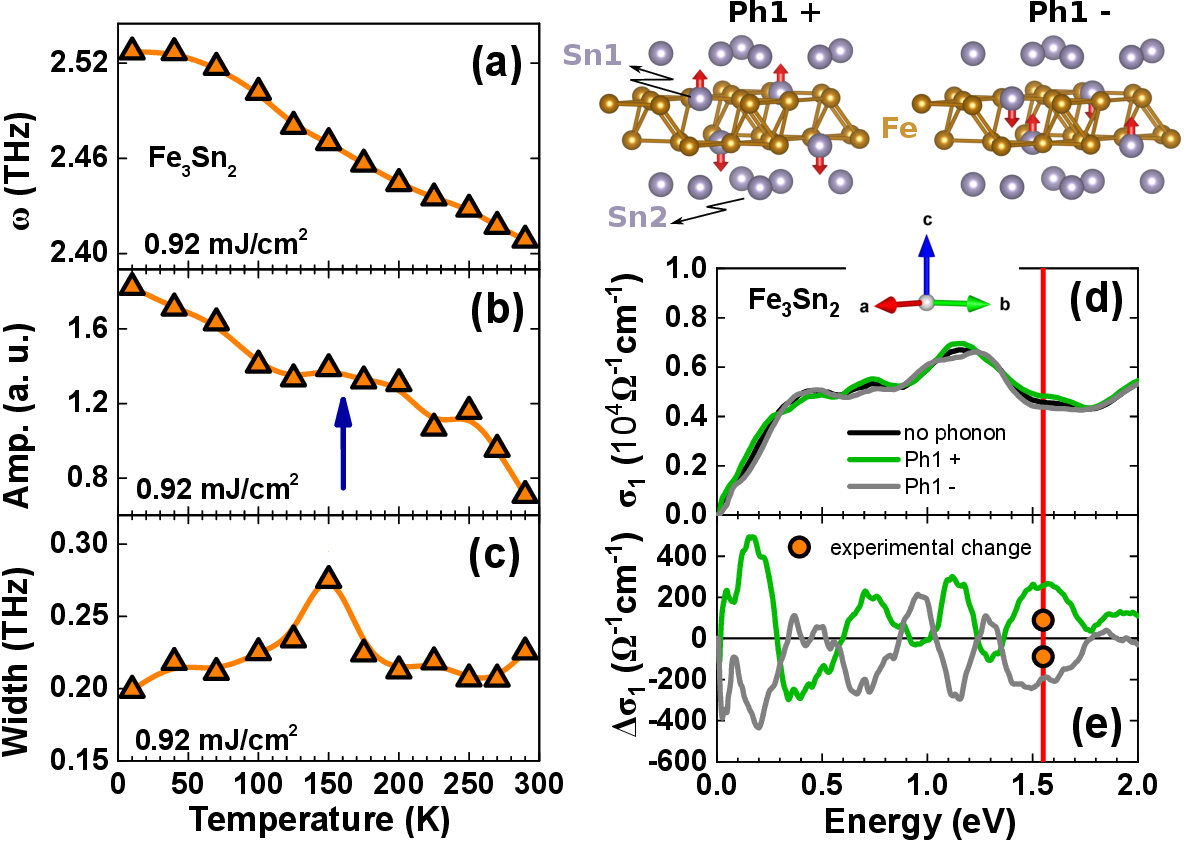}
      \caption{Temperature dependence of (a) frequency, (b) amplitude, and (c) width of the A$_{1g}$ phonon mode in \FeSn. Arrow in (b) is to highlight the temperature range where the system goes through a spin reorientation. (d) presents the calculated optical conductivtiy via DFT for the nominal \FeSn\ along with the distorted structure under A$_{1g}$ phonon mode. (e) depicts the change in optical conductivity under the influence of the phonon mode. The orange circles are the change of the optical conductivity estimated by changing the experimental reflectivity~\cite{Biswas2020Aug} by 10$^{-3}$. The red line in (d) and (e) is the pump-probe frequency of 800 nm. Bilayer kagome structure have been given with the representation of the A$_{1g}$ phonon mode demonstrated with the red arrows.}
    \label{F3}
\end{figure}

Fig.~\ref{F3}(a-c) depicts the temperature dependence of the obtained phonon parameters, namely the resonance frequency, amplitude and width. Its frequency, retrieved using a Fourier transform was found to be around $2.40$ to $2.50$~THz, which corresponds to an A$_{1g}$ totally symmetric mode~ \cite{PhononModes}. As shown in Fig.~\ref{F3}, this is primarily an out-of-plane Sn mode that does not affect the kagome network significantly. It is dependent on both temperature and magnetic structure, presenting a clear phonon softening with increasing temperature and anomalies on its amplitude and peak width around the spin reorientation region $\sim150$ K, as demonstrated with the blue arrow in Fig.~\ref{F3}(b). The phonon softening and the increase of the amplitude of the phonon oscillations have also been observed with increasing fluence [See suplementary Fig.S5].  

Phonon softening with temperature and fluence is often attributed to anharmonic terms in the vibrational potential energy \cite{Anharmonic_effects_light_scattering, Misochko2004May}. However, other signatures of these anharmonic effects are not observed in our data. For instance, the increase of the amplitude does not follow the expected increasing behavior. Furthermore, the width does not show a decrease, in fact a slight increase at higher temperatures indicates a strong electron-phonon coupling. Other evidence against the anharmonic phonon softening is that the decay rate of the phonon does not change significantly with temperature (details of the analysis are given in supplementary material). 

Along with the evidences against the anharmonic phonon coupling, the absence of $E_g$ phonon modes, the cosine-like character of the oscillations [see supplementary], and the large amplitude of the oscillations when compared to the non-oscillatory decaying signal (also increasing linearly with fluence), are strong indications of displacive excitation of coherent phonons (DECP) as the mechanism behind this coherent phonon generation \cite{DECP_Theory}. This indicates a strong electron-phonon coupling $(e-p)$ in \FeSn\ in both low and room temperature regimes, as DECP depends exclusively on this coupling to induce the coherent oscillations.  The maximum of the non-oscillatory exponential decay increases with fluence, indicating a larger photo-excited carriers density at higher fluences, and then a considerable electronic softening of the lattice is expected \cite{diamond_instability,softening_tellurium}. As a consequence, the reduction of the restoring force for the A$_{1g}$ lattice displacement appears naturally with the excitation of a larger number of electrons. Thus, this phenomenon can be understood as solely an electronic softening of the crystal lattice.

Such a strong phonon coupling suggests some sort of an incipient lattice distortion in \FeSn. In first glance, the breathing kagome distortion [Fig.~\ref{F1}(b)], where the successive Fe-bonds in kagome network are slightly different, is a reasonable cause. On the other hand, in this case, it is expected that the breathing $E_g$ mode, which directly affects the kagome network, would be the phonon that modulates reflectivity. Considering that the observed A$_{1g}$ mode does not affect this breathing kagome structure, this assumption seems to be unlikely. 

Another possibility why \FeSn\ is special lies in the proclivity of kagome metals for charge-density-wave instabilities that have been revealed not only in nonmagnetic compounds like AV$_3$Sb$_5$ and ScV$_6$Sn$_6$ \cite{Jiang2021Oct, Arachchige2022Nov, Ratcliff2021Nov, Wang2021Oct}, but also in the magnetic kagome metal FeGe \cite{Teng2023Jun}. Our data support growing evidence that even in the absence of a CDW transition, charge carriers in kagome metals can be strongly coupled to specific phonons that, in turn, have crucial effect on their dynamics.. 

Finally, we use density-functional-theory (DFT) to elucidate the effect of the A$_{1g}$ phonon mode on the optical conductivity, details of the calculations are given in supplementary. We have introduced the atomic displacements due to the phonon mode as demonstrated in Fig.~\ref{F3}, and calculated the optical conductivity as given in Fig.~\ref{F3}(d). The displacement amplitude is taken as 0.1~\r A that is consistent with the estimated atomic displacement (see supplementary). To demonstrate changes in the optical conductivity, and ensuing changes in the reflectivity, we have plotted in Fig.~\ref{F3}(e) the difference in optical conductivity with respect to the undistorted structure. The results suggest that at $800$~nm [red line in Fig.~\ref{F3}(e)], the observed 2.5 THz phonon mode has a strong impact on the optical conductivity and can clearly be the reason behind the observed 10$^{-3}$ change in the reflectivity (the orange circles are the estimates over the experimental reflectivity spectra). The distortion of the structure in two opposite directions nicely leads to a symmetric change of the optical conductivity. For comparison, changes in the optical conductivity induced by the other A$_{1g}$ modes (presented in the supplementary materials) have also been calculated. Results suggest that at $800$~nm, the most prominent change is due to the observed 2.5 THz A$_{1g}$ mode, and other modes do not alter the optical conductivity significantly. These calculations may also explain why we could measure only a single phonon mode as a reflectivity modulation, while the other totally symmetric A$_{1g}$ modes have not be observed.        

\textcolor{blue}{\textit{Conclusions.}} Photo-induced changes in reflectivity of the kagome metal \FeSn\ reveal the dynamics of carriers and coherent optical phonons. We detect three time scales. 
Two of them, the faster and slower ones, are clearly related to the highly metallic nature of the material and can be well explained using the two-temperature model for metals. The medium time scale, on the other hand, is related to the unconventional localized carriers in kagome metals. Their distinct relaxation time and coupling to short optical pulses allows an independent probe of Drude and localized carriers, as well as the control of localization using ultrafast optical probes. Additionally, strong coherent phonon oscillations have been observed indicating a strong electron-phonon coupling in \FeSn\ even at room temperature. The nature of this phonon mode is attributed to the electronic softening of the crystal lattice due to the large photo-induced carrier density. The spin reorientation of \FeSn\ around $150$~K does not seem to have a significant effect on the dynamics of charge carriers, although it manifests itself in the temperature dependence of the coherent phonon. In conclusion, our study demonstrates the salient role of phonon dynamics and electron-phonon coupling even in those kagome metals where no CDW instabilities occur.

\begin{acknowledgments}
H. C. L. acknowledges support from the National Key R\& D Program of China (Grants
No. 2016YFA0300504 and No. 2018YFE0202600), and the National Natural Science Foundation of China
(Grants No. 11574394, No. 11774423, and No. 11822412). The work in Germany has been supported by the Deutsche Forschungsgemeinschaft (DFG) via Grant UY63/2-1. Computations for this work were done (in part) using resources of the Leipzig University Computing Center.
\end{acknowledgments}

\bibliography{Fe3Sn2_references}

\makeatletter
\renewcommand\@bibitem[1]{\item\if@filesw \immediate\write\@auxout
    {\string\bibcite{#1}{S\the\value{\@listctr}}}\fi\ignorespaces}
\def\@biblabel#1{[S#1]}
\makeatother
\setcounter{figure}{0}
\renewcommand{\thefigure}{S\arabic{figure}}
\newcommand{\av}{\mathbf a}
\newcommand{\bv}{\mathbf b}
\newcommand{\cv}{\mathbf c}

\newpage
\begin{center}
\textbf{Supplementary Material for\\ ``Ultrafast Carrier and Phonon Dynamics of the Magnetic Kagome Metal \FeSn"\\}
\end{center}

\begin{center}
M. V. Gonçalves-Faria, A. Pashkin, Q. Wang, H. C. Lei, S. Winnerl, A. A. Tsirlin, M. Helm, and E. Uykur
\end{center}

\subsection{Samples and the Experimental Details}
Single crystals of \FeSn\ were grown using self-flux method as described elsewhere \hyperlink{Wang2016}{\color{blue}[1]}. The (001)-plane with lateral dimensions of 1000~$\mu$m$ ~\times~ $800~$\mu$m$ ~\times$ 200~$\mu$m is used for the optical pump-probe transient reflectivity measurements. We used the same as-grow sample as in our previous infrared spectroscopy study \hyperlink{Biswas2020}{\color{blue}[2]}.

Temperature- and fluence- dependent optical pump-probe spectroscopy were performed in the reflection geometry. For pump and probe we used 60~fs long laser pulses, centered at 800~nm and generated by a Ti:sapphire laser amplifier with 250 kHz repetition rate. Probe spot on the sample was around 25~$\mu$m and the pump spot was around 30~$\mu$m. 

\subsection{Data Analysis} 

Transient reflectivity was measured up to 150~ps delay time with 1~ps time resolution. Up to around 9~ps delay time, we increased the time resolution to 33 fs to resolve the phonon oscillations. The overall temperature and fluence dependence of the transient reflectivity does not change drastically and can be analysed by employing equation (1) from the main text. $\tau_1$ and $\tau_2$ are the retrieved relaxation times, and $y0$, the offset parameter, is also related to a relaxation process, as described in the main text; however, since it is much longer than our measurement limit, the relaxation time cannot be obtained reliably. Therefore, we model it with a constant.

\begin{figure}
    \centering
        \includegraphics[width=\columnwidth]{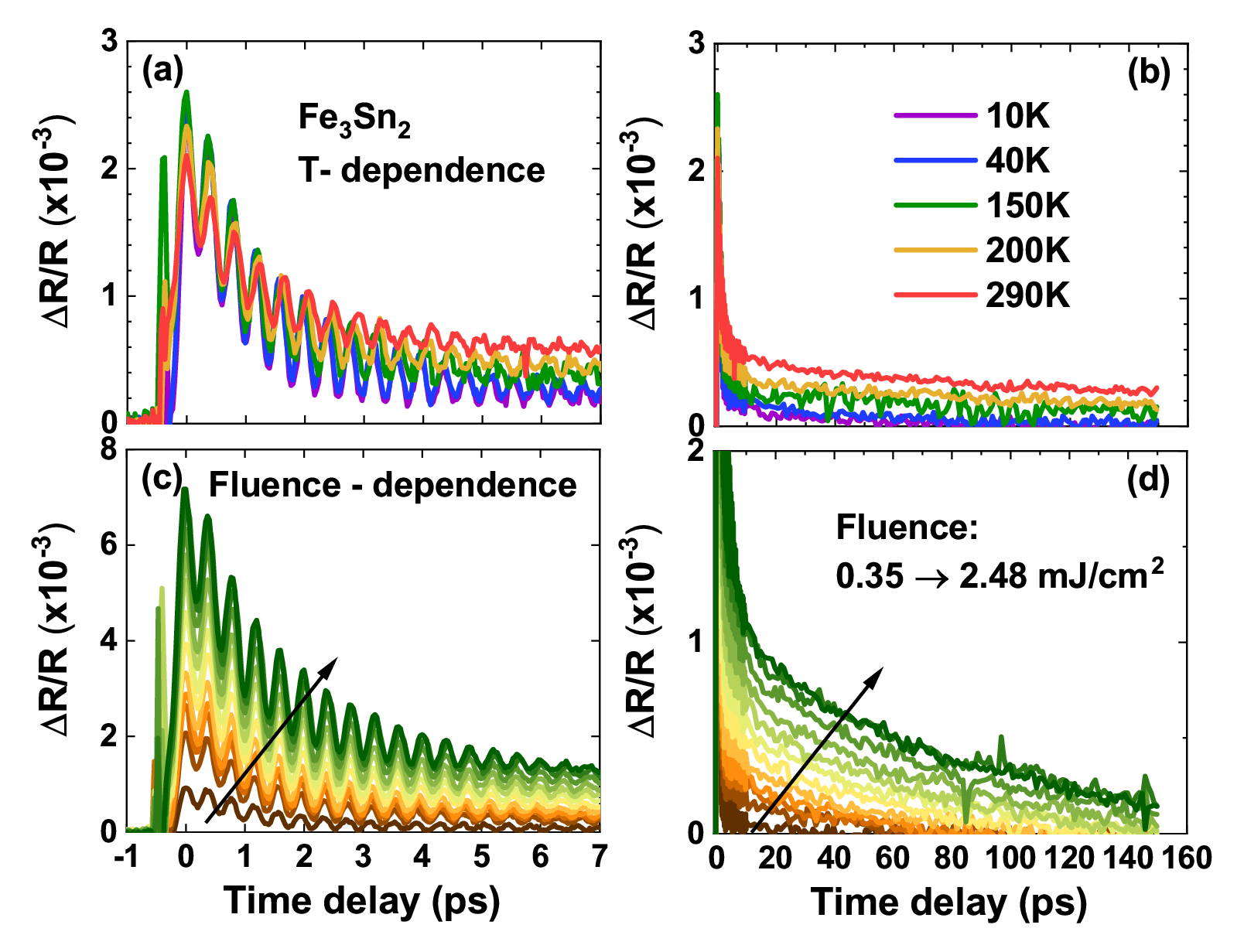}
    \caption{(a, b)Temperature and (c,d) 10 K fluence dependence of the transient reflectivity in the short and long time delays. }
    \label{F1s}
\end{figure}

In Fig.~\ref{F1s}, the temperature and fluence dependence of the measured transient reflectivity can be seen in the short and long time delays. The oscillations dominating the short time delays are the coherent phonons. The analysis process of these are given below and the details are discussed in the main text. 

\subsection{Phonon Mode} 

\begin{figure}
    \centering
        \includegraphics[width=\columnwidth]{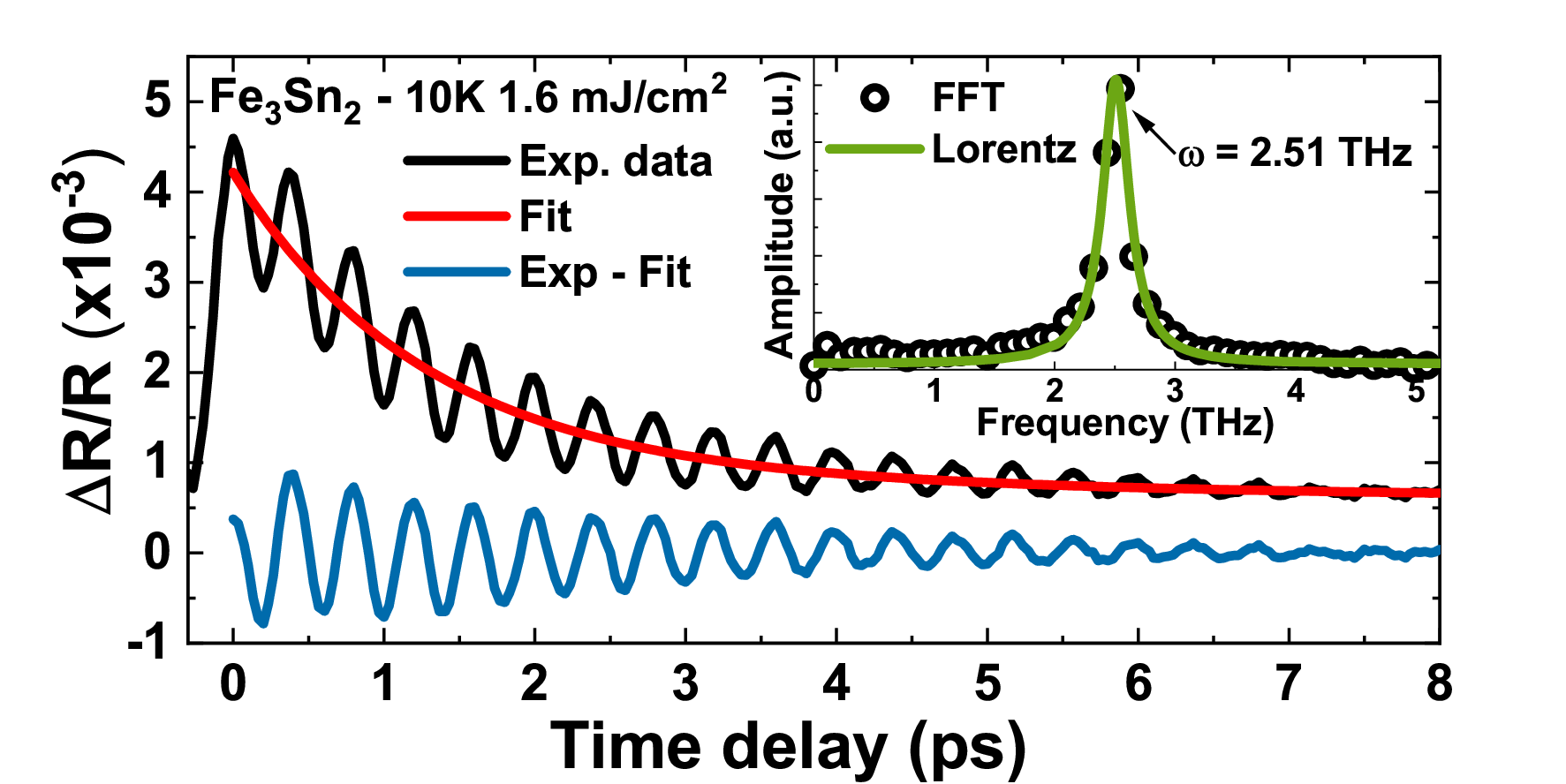}
    \caption{Coherent optical phonon analysis for the \FeSn\ pump-probe signal at $10~$K and 1.6~mJ/cm$^2$. Experimental data and the exponential fit are the black and red solid lines, respectively. The blue line is the result of subtracting the fit from the experimental data. The inset shows the FFT of the oscillatory blue line with a Lorentzian fit to find the resonance frequency, which is 2.51~THz in this case.}
    \label{F2s}
\end{figure}

The short delay time of the transient reflectivity is dominated by the coherent phonon oscillations. Here, we fit the non oscillatory part of the transient reflectivity as demonstrated in Fig.~\ref{F2s}. Afterwards, this fit is subtracted from the main signal to obtain the isolated oscillations as shown with the blue curve in the same figure. The fast Fourier transform (FFT) of these oscillations gives the resonance frequency of the phonon mode that couples to the electronic background. As depicted in the inset of Fig.~\ref{F2s}, we fitted the FFT with a Lorentzian in order to obtain the resonance frequency, amplitude, and the width of the phonon mode, as discussed in the main text.

\begin{figure}
    \centering
        \includegraphics[width=1\columnwidth]{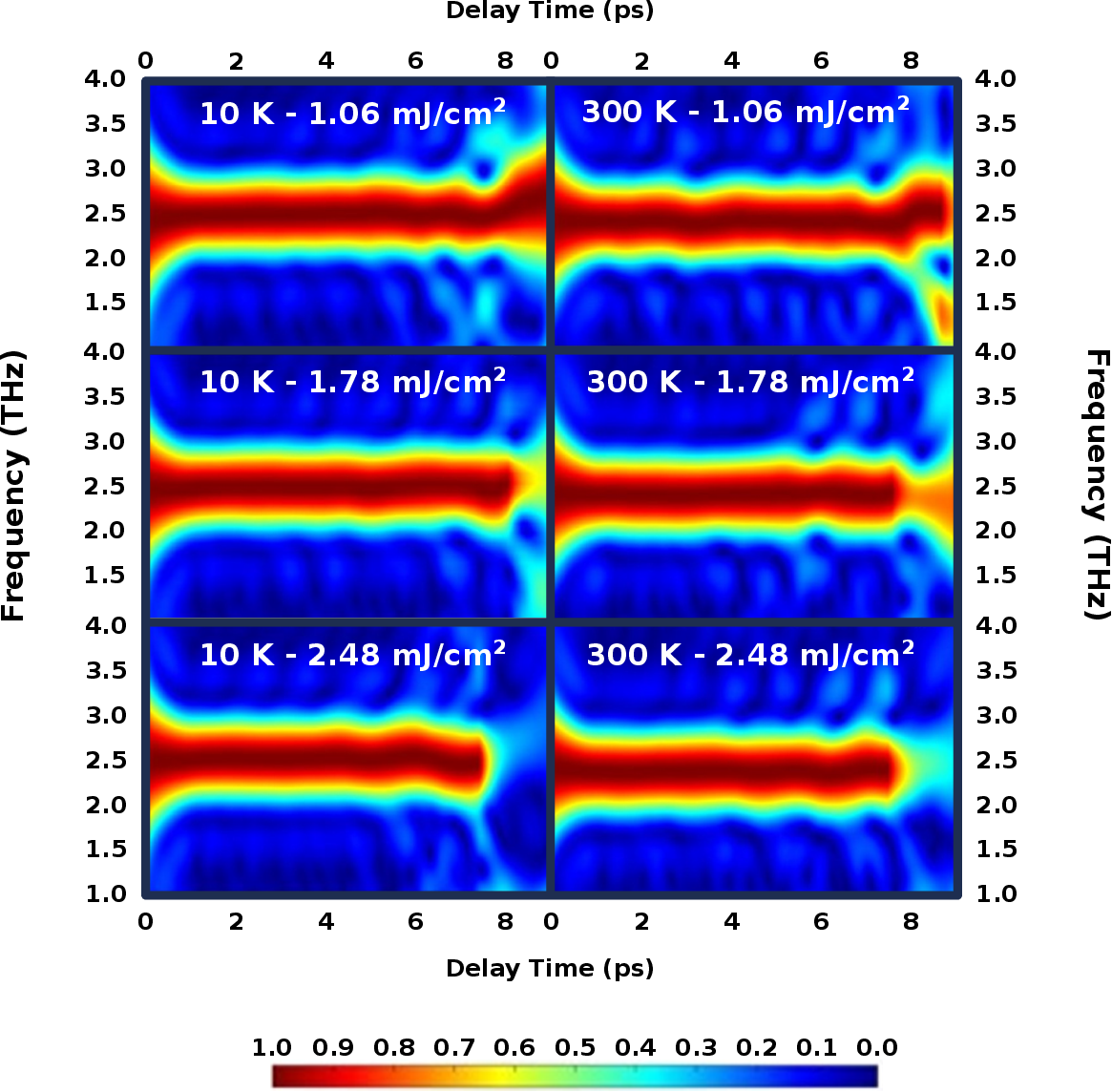}
      \caption{Temperature- and fluence dependence of the wavelet transform of the frequency of the oscillations. $A_{1g}$ is the only phonon mode observed and the frequency does not change with increasing time delay. The decay rate of the phonon mode is increasing with fluence; however, does not change significantly with temperature.}
    \label{F3s}
\end{figure}

The details of the observed coherent phonon oscillations were analyzed with the wavelet transform of the frequency of the oscillations as given in Fig.~\ref{F3s}. The decay rate of the phonon does not change significantly with temperature. It is around 0.32~ps$^{-1}$ both at room temperature and 10~K. With increasing fluence, it slightly increases to 0.35~ps$^{-1}$, nonetheless, does not change appreciably with temperature. Moreover, here it is demonstrated that the $A_{1g}$ mode is the only mode observed and it does not vary with time.

\begin{figure}
    \centering
    \includegraphics[width=0.8\columnwidth]{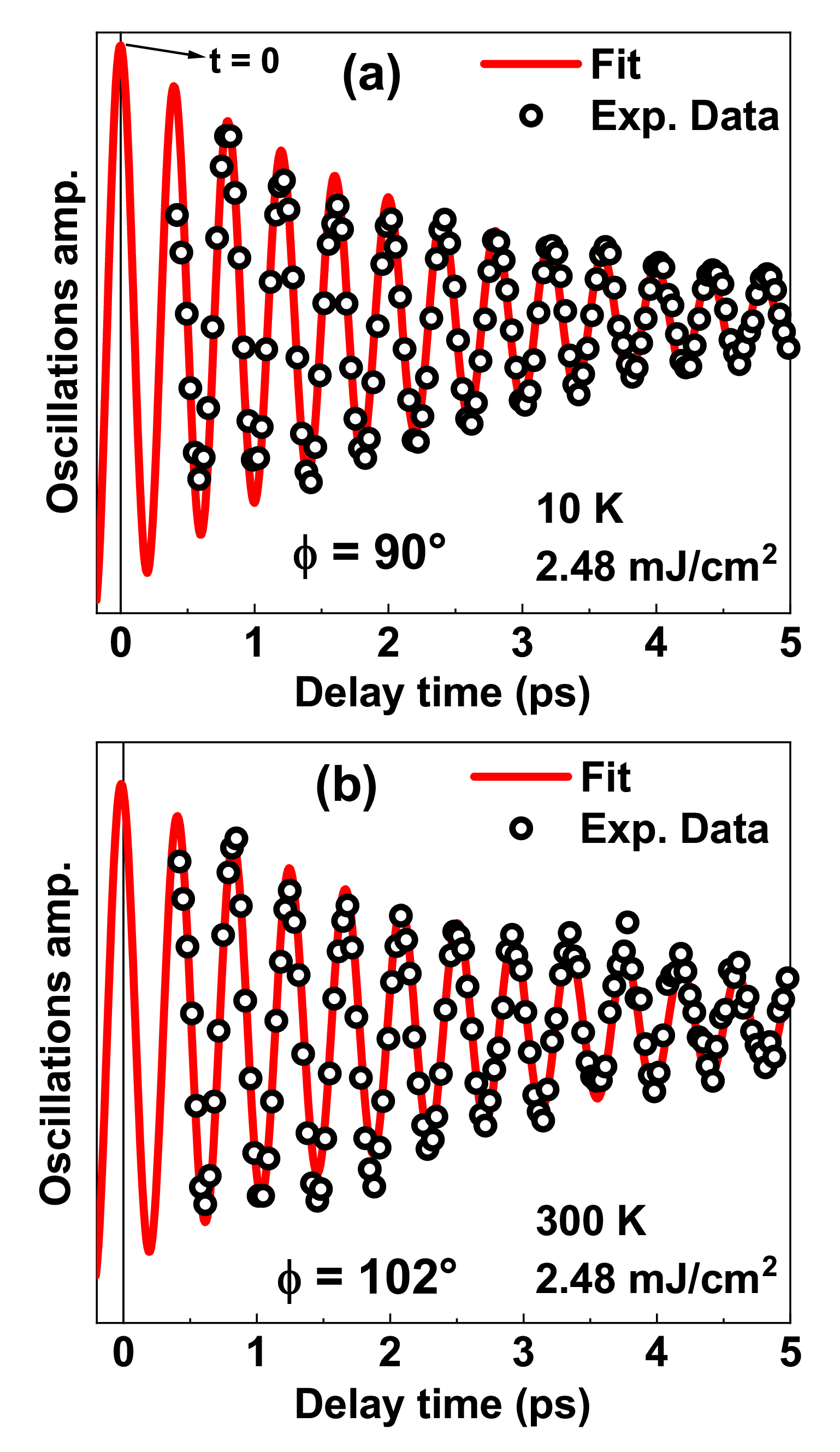}
    \caption{Fit for the coherent oscillations at (a) 10 K and (b) 300 K using Eq.~\ref{E7}.}
    \label{F7}
\end{figure}

As depicted in Fig.~\ref{F3s}, we observed no chirp in the $A_{1g}$ phonon mode, in contrast to what has been observed in other compounds, such as Bismuth for example \hyperlink{Misochko2004}{\color{blue}[3]}. Therefore, we can fit the oscillatory part of the signal (blue line in Fig.~\ref{F2s}) with the following equation
\begin{equation}
    \centering
    y = Ae^{-(t/t_0)}\text{sin}(\omega t + \phi),
    \label{E7}
\end{equation}
in order to retrieve its phase $\phi$. This is an additional way of testing the displacive excitation of coherent phonons (DECP) as the generating mechanism of the coherent phonons, since the oscillations should present a cosine-like behavior \hyperlink{Zeiger1992}{\color{blue}[4]}. Fig.~\ref{F7} displays the experimental data (dots) and the fitting (red solid line) using Eq.~\ref{E7} at 10 K (panel a) and room temperature (panel b). The oscillations were extrapolated until time zero, that was determined from the pump-probe interference pattern generated before the increase in reflectivity. At 10 K the phase is 90° away from a sine function and at 300 K the phase is 102°, which indicates a cosine description, being; therefore, in good agreement with a DECP-launched coherent phonon.

\begin{figure}
    \centering
        \includegraphics[width=1\columnwidth]{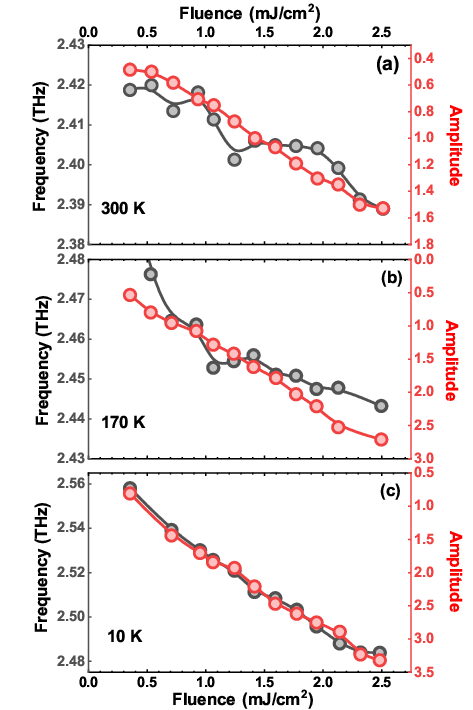}
      \caption{Frequency (grey symbols) and the amplitude (red symbols) of the phonon mode for 300~K (a), 170~K (b), and 10~K (c). Plots are scaled to each other to demonstrate the mutual change. }
    \label{F4}
\end{figure}

The fluence dependence of the phonon is also investigated. In the main text, the phonon softening with increasing fluence for different temperatures has been discussed along with the changes on its amplitude (increasing with fluence). Here we plot (Fig.~\ref{F4}) the resonance frequency and the amplitude on the same graph demonstrating the remarkable match between them as another evidence to the DECP mechanism. Please note that lower fluences, especially at higher temperatures, show a worse signal to noise ratio, hence some deviations are observed. \\

\subsection{Heat Capacity and Two Temperature Model}

\begin{figure}
    \centering
        \includegraphics[width=1\columnwidth]{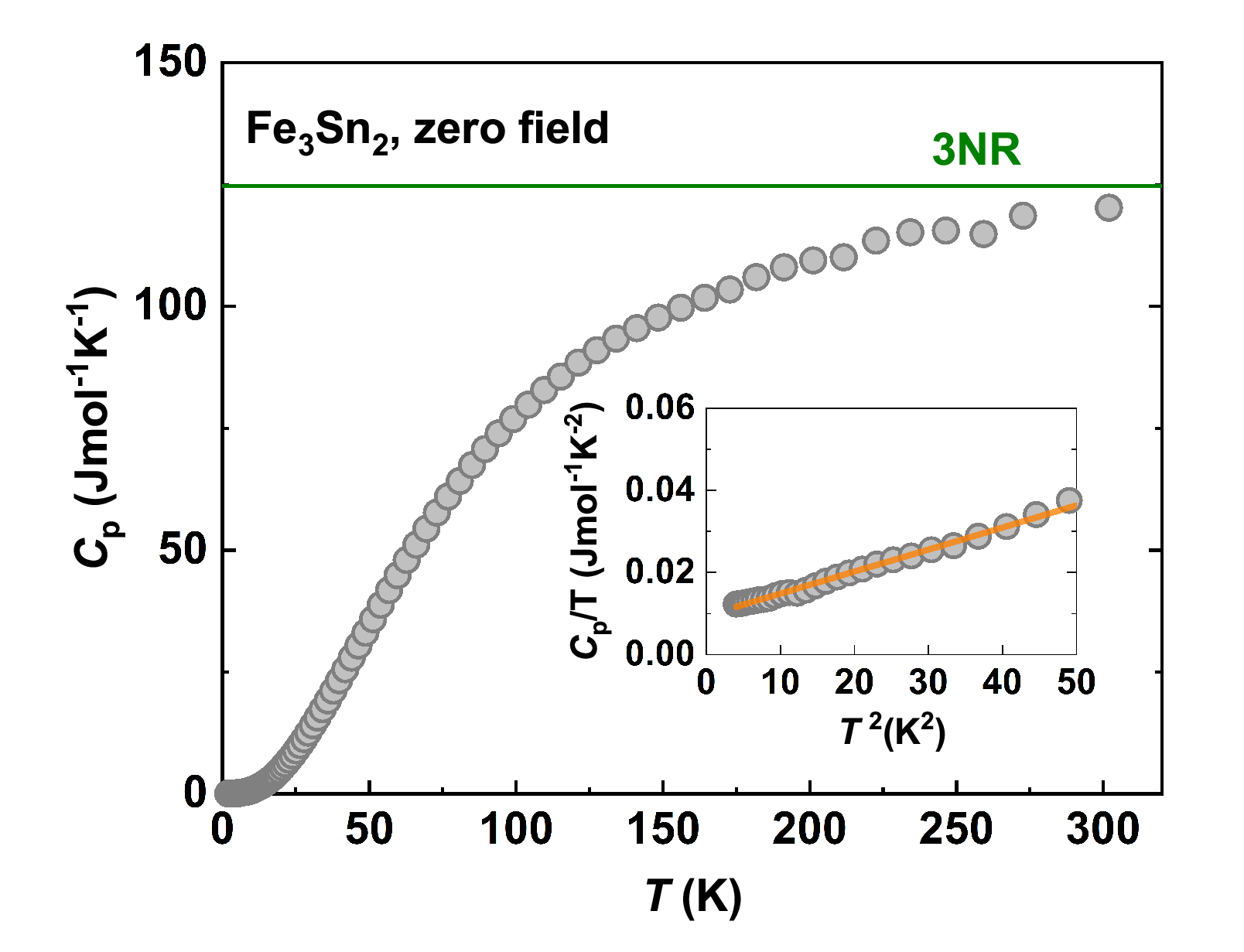}
      \caption{Temperature dependence of the heat capacity. The inset shows C$_p$/T vs T$^2$ and its linear fitting.}
    \label{F5}
\end{figure}

In Fig.~\ref{F5}, we plotted the temperature dependence of the heat capacity between 2 and 300~K provided by the crystal grower (H. Lei unpublished). The smooth evolution of the heat capacity confirms the absence of structural phase transitions in this temperature range. At high temperatures, the heat capacity reaches the Dulong-Petit limit. The inset demonstrates the C$_p$/T vs T$^2$ with its linear fit according to:
\begin{equation}
    \centering
    C_p = \gamma T + \frac{12}{5}\pi ^4nR\left(\frac{T}{\Theta_D}\right)^3,
    \label{E3}
\end{equation}
where $\gamma, n, R,$ and $\Theta_D$ are the Sommerfeld coefficient of the electronic contribution, the number of atoms per formula unit, molar gas constant, and the Debye temperature of the lattice contribution, respectively. Our fit gives $\gamma=$ 9.45~mJmol$^{-1}$K$^{-2}$ and $\Theta_D=$ 263 K.
  
\begin{figure}
    \centering
    \includegraphics[width=1\columnwidth]{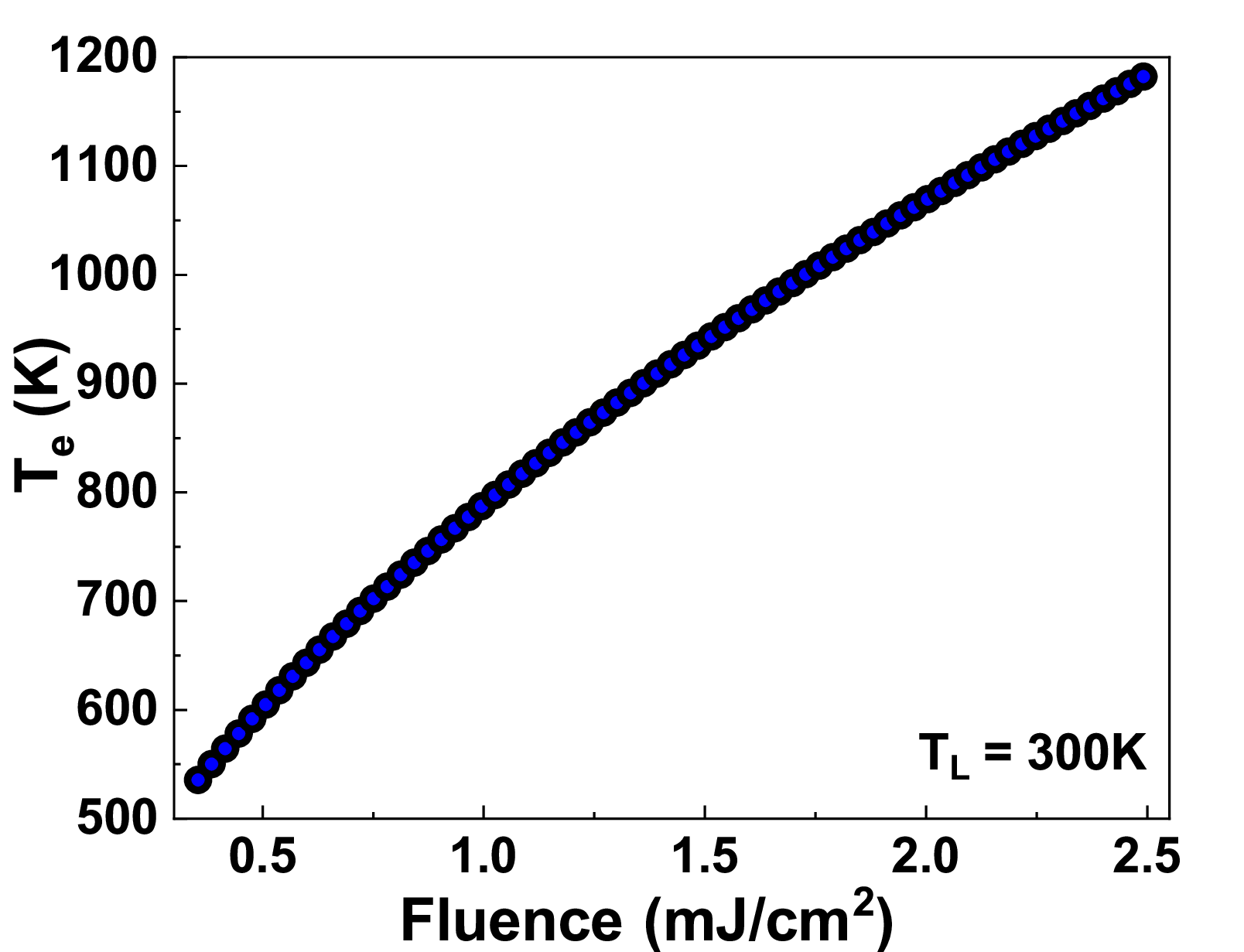}
    \caption{Electronic temperature  $T_e$ as function of fluence according to Eq.~\ref{E5} with $T_l = 300K$.}
    \label{F8}
\end{figure}

\begin{figure}
    \centering
    \includegraphics[width=1\columnwidth]{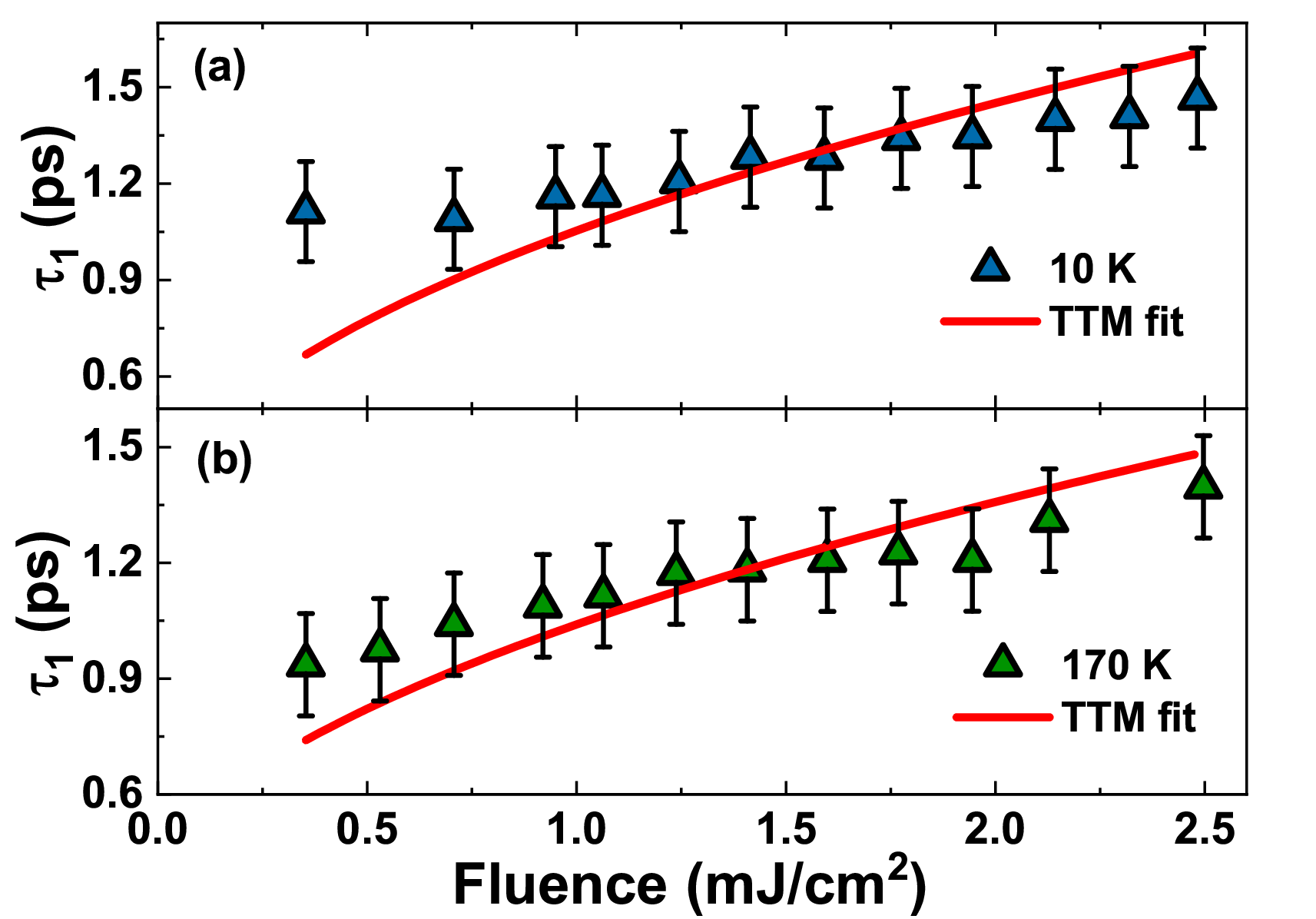}
    \caption{Fluence dependence of $\tau_1$ at (a) 10K and (b) 170K. Red solid lines are fits using Eq.(\ref{E4}) from the two-temperature model.}
    \label{F6}
\end{figure}

The parameters $\tau_1$ and $y_0$, retrieved from the exponential fitting of the experimental data, can be explained by the two-temperature model (TTM) \hyperlink{Ultrafast_metals2}{\color{blue}[5]}. In this model, the metal is modelled as a two coupled thermal systems composed of electrons and the crystal lattice, respectively. A difference between electronic and lattice temperatures can be induced by the incident ultrashort laser pulse, by transferring its energy to the conduction electrons which thermalize rapidly via electron-electron scattering. Since the electronic heat capacity C$_e$ is much less than the lattice heat capacity C$_l$, it is possible to create transient electronic temperatures much higher than the lattice temperature. Then in a time scale of the order of $\tau_1$, these hot electrons return to a local equilibrium solely through electron-phonon scattering processes \hyperlink{Ultrafast_metals}{\color{blue}[6]}. In such systems a much longer relaxation process is also observed. This results from residual lattice heating which cools down in a longer time scale via heat diffusion. Based on the TTM, the relaxation time associated with the electron-phonon scattering, $\tau_1$, is given by \hyperlink{Groeneveld1995}{\color{blue}[7]}:
\begin{equation}
    \tau = \frac{\gamma(T_{e}^2 - T_{l}^2)}{2H(T_{e},T_{l})},
    \label{E4}
\end{equation}
where $\gamma$ is the Sommerfeld coefficient calculated with the heat capacity data from Fig. \ref{F5} and fitted with Eq.(\ref{E3}), $T_e$ and $T_l$ are the temperatures of the electronic and lattice subsystems, respectively, and $H(T_e,T_l)$ is the energy transfer rate per unit volume and per second from electrons to phonons, given by:
\begin{equation}
    H(T_e,T_l) = f(T_e) - f(T_l),
\end{equation}
where
\begin{equation}
    f(T) = 4G_{\infty}\frac{T^5}{\Theta_D^4}\int_{0}^{\Theta_D/T}\frac{x^4}{e^x -1}dx,
\end{equation}
with $G_\infty$ being the electron-phonon coupling constant. The electronic temperature can be estimated by
\begin{equation}
    T_e = \left(T_l^2 + \frac{2U_l}{\gamma}\right)^{1/2},
    \label{E5}
\end{equation}
where $U_l$ is the deposited laser energy density calculated using the penetration depth of \FeSn\ and the incident fluences. The fluence dependence of $T_e$ at room temperature according to Eq.~\ref{E5} is shown in Fig.~\ref{F8}. Figure 2(g) of the main text presents the fit for $\tau_1$ as function of fluence at room temperature using Eq.(\ref{E4}). At $10$ and $170$~K the results and the fits for $\tau_1$ are presented in Fig. \ref{F6}. In all cases an increase of $\tau_1$ with encreasing fluence was observed, which is the behavior expected from the TTM.

\begin{table}[h!]
    \centering
    \caption{Electron-phonon coupling constant as function of temperature.}
    \begin{tabular}{c|c}
    \textbf{T(K)} & \textbf{G$_\infty$(Wm$^{-3}$K$^{-1}$)$\times10^{17}$} \\
    \hline
    10 &  1.50 \\
    170 & 1.23 \\
    300 & 1.17 \\
    \end{tabular}
    \label{T1}
\end{table}

In Table~(\ref{T1}) the retrieved values for the electron-phonon coupling constant G$_\infty$ are summarised. These are the results from the best fits of the experimental data to the model in Eq.~(\ref{E4}). One can see that G$_\infty$ decreases with increasing temperature, which is in agreement with the phonon amplitude shown in Fig.(3b) of the main text. 

\subsection{Density Functional Theory Calculations}
Density-functional-theory (DFT) calculations of the band structure were performed in the \texttt{Wien2K}~\hyperlink{wien2k}{\color{blue}[8,9]} code using Perdew-Burke-Ernzerhof flavor of the exchange-correlation potential~\hyperlink{pbe96}{\color{blue}[10]}. We used experimental structural parameters determined by x-ray diffraction measurements, as summarized in Table~(\ref{structure}). The \texttt{optic} module~\hyperlink{optic}{\color{blue}[11]} was used for evaluating the optical conductivity. Spin-orbit coupling was included in the calculations of band structure and optical conductivity. Furthermore, ferromagnetic order was taken into account, where the spins on Fe-atoms are aligned along in-plane direction, which eventually occurs below the spin-reorientation temperature. The DFT-obtained magnetic moment per Fe-atom was 2.18~$\mu_B$, which is very close to the experimental values \hyperlink{Wang2016}{\color{blue}[1]}.

\begin{table}[h!]
\centering
\caption{Details of x-ray diffraction data collection and refined structural parameters for \FeSn.}
\begin{tabular}{c}\hline
 $a=b=5.3311(6)$\,\r A,\quad $c=19.7644(35)$\,\r A \\
 $V=484.46$\,\r A$^3$ \\
 $R\bar{3}m$ \\
 $\lambda=0.71073$\,\r A \\
 $2\theta_{\max}=65.37^{\circ}$ \\
 $-8\leq h\leq 7$,\quad $-6\leq k\leq 7$,\quad $-25\leq l\leq 29$ \\
 $R_{I>4\sigma(I)}=0.0652$ \smallskip\\\hline
\end{tabular} 
\\\medskip
\begin{tabular}{c@{\hspace{0.5cm}}c@{\hspace{0.5cm}}c@{\hspace{0.5cm}}c}
 Parameter & Fe & Sn1 & Sn2 \\
 $x/a$ & 0.49595(31) & 0 & 0 \\
 $y/b$ & 0.50405(31) & 0 & 0 \\
 $z/c$ & 0.11330(17)  & 0.10430(14)   & 0.33147(13) \\
 $U_{\rm iso}$ (\r A$^2$) & 0.03498(76) & 0.03639(70) & 0.03543(68)
\smallskip\\\hline
\end{tabular}
\label{structure}
\end{table}

Self-consistent calculations were converged on the $24\times 24\times 24$ $k$-mesh. Optical conductivity was calculated on the $k$-mesh with up to $100\times 100\times 100$ points within the Brillouin zone.

\begin{table*}
\centering
\begin{small}
\caption{Calculated Gamma-point phonon frequencies and their symmetries.}
 \begin{tabular}{c@{\hspace{0.5cm}}c@{\hspace{0.5cm}}c@{\hspace{0.5cm}}c@{\hspace{0.5cm}}c@{\hspace{0.5cm}}c}
    \hline
    Phonon mode (\cm) & Symmetry & IR/Raman & Phonon mode (\cm) & Symmetry & IR/Raman\\ \hline
    259.6 & E$_g$ & Raman & 178.7 & A$_{2u}$ & IR \\ \hline
    253.7 & E$_u$ & IR & 152.0 & E$_u$ & IR\\ \hline
	230.0 & A$_{1g}$ \textbf{(Ph4)} & Raman & 143.7 & E$_g$ & Raman\\ \hline
	228.9 & A$_{1u}$ & - & 136.6  & A$_{1g}$ \textbf{(Ph2)} & Raman\\ \hline
	224.4 & A$_{1g}$ \textbf{(Ph3)} & Raman & 135.4 & E$_g$ & Raman\\ \hline
	211.4 & A$_{2u}$ & IR & 114.7 & A$_{2u}$ & IR\\ \hline
	203.4 & E$_u$ & IR & 92.4 & E$_u$ & IR\\ \hline
	196.2 & E$_g$ & Raman & 90.2 & E$_g$ & Raman \\ \hline
	189.5 & A$_{2g}$ & - & 87.8  & A$_{1g}$ \textbf{(Ph1)} & Raman \\ 
	    \hline
  \end{tabular}
\label{phonon} 
\end{small}
\end{table*}

\begin{figure*}
    \centering
        \includegraphics[width=2\columnwidth]{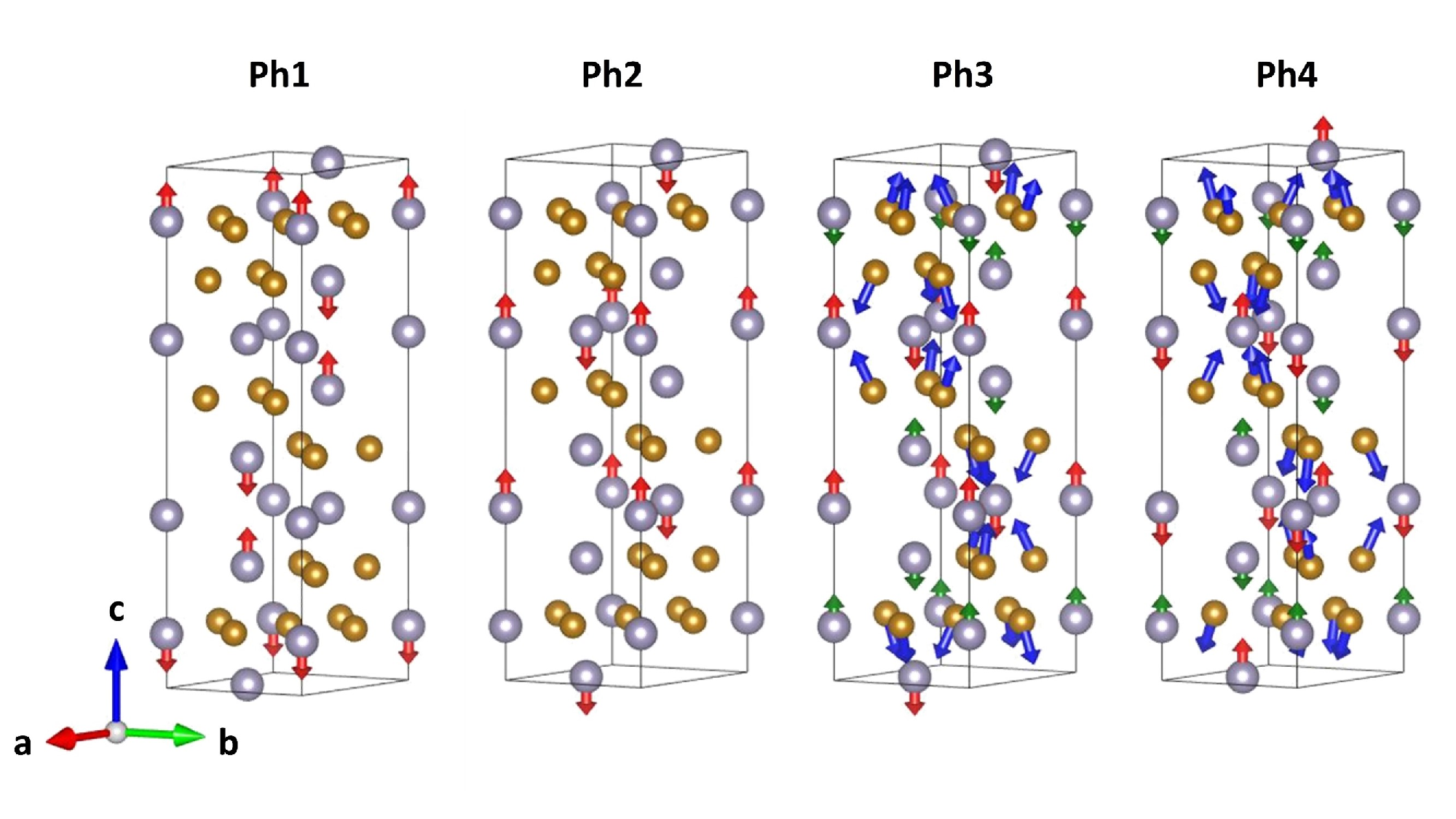}
      \caption{Representations of $A_{1g}$ totally symmetric phonon modes. Grey circles are Sn atoms and yellow circles are Fe atoms.}
    \label{Ph}
\end{figure*}

The phonon calculations were performed in \texttt{VASP} using the same structural parameters and the built-in procedure with frozen atomic displacements of 0.015~\AA. Magnetic moments were directed along the $c$ axis to avoid symmetry lowering. The obtained Gamma-point phonon energies are given in Table~(\ref{phonon}). Relevant to the current work, the four $A_{1g}$ modes are also depicted in Fig.~\ref{Ph}, where the Ph1 is the coherent phonon oscillations in this pump-probe study. The conductivity change under the influence of phonon modes was investigated with DFT by deforming the structure under the mentioned phonon mode. The procedure is described below.  

\begin{figure}
    \centering
        \includegraphics[width=1\columnwidth]{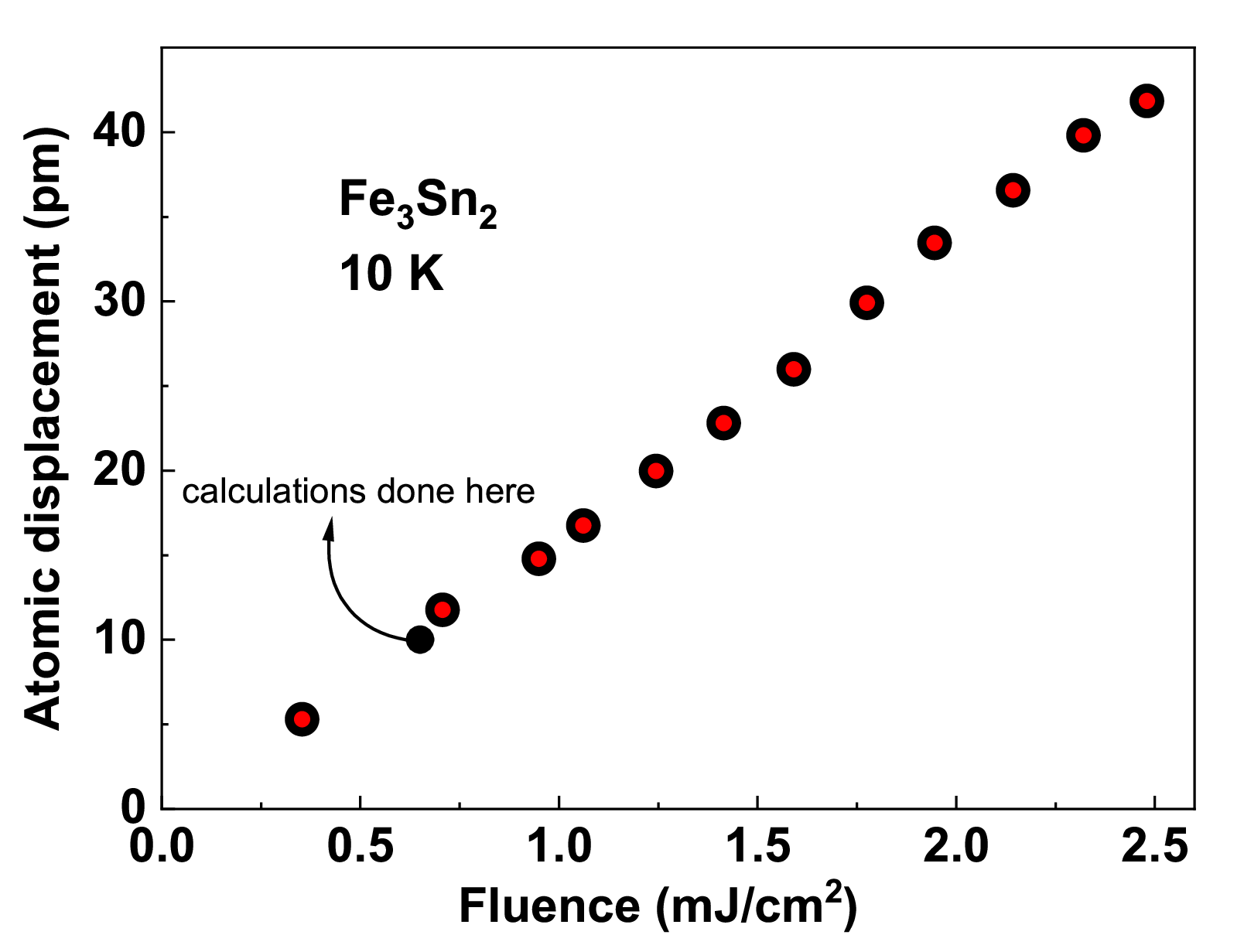}
      \caption{Experimental displacement amplitude as a function of fluence at 10~K. Black circle shows the value used in the DFT calculations.}
    \label{AD}
\end{figure}

Firstly, we estimated the amplitude of the atomic displacement for the relevant Ph1 $A_{1g}$ using our transient reflectivity experimental data. Calculating the exact values of coherent displacement amplitude in ultrafast structural dynamics is a great challenge. However, the order of magnitude of such atomic displacements may be determined from the reflectivity variation $\Delta$R/R using the following equation, that is based on equations (11) and (12) from Ref.~\hyperlink{Stevens2002}{\color{blue}[12]}:
\begin{equation}
\frac{1}{R}\frac{\partial R}{\partial E_c}Q_0^2 = \frac{\Delta R}{R}\frac{Im(\varepsilon)}{h\Omega_0^2\rho c}\frac{\Phi}{1-R},
\label{LatDis}
\end{equation}
where $Q_0$ is the displacement amplitude, $E_c$ is the incident laser pulse frequency, $\Omega_0$ is the coherent phonon frequency, $\rho$ is the mass density of the oscillating atoms (Sn in this case) in the material, $c$ is the speed of light, $\Phi$ is the incident fluence on the surface, $h$ is Planck's constant, and $\varepsilon$ is the dielectric constant. The measured values of $\Delta$R/R around 10$^{-3}$ result in $Q_0$ of about 5-40~pm. The fluence dependence for the displacement amplitude according to Eq.~(\ref{LatDis}) is given in Fig.~\ref{AD}.

\begin{figure}
    \centering
        \includegraphics[width=1\columnwidth]{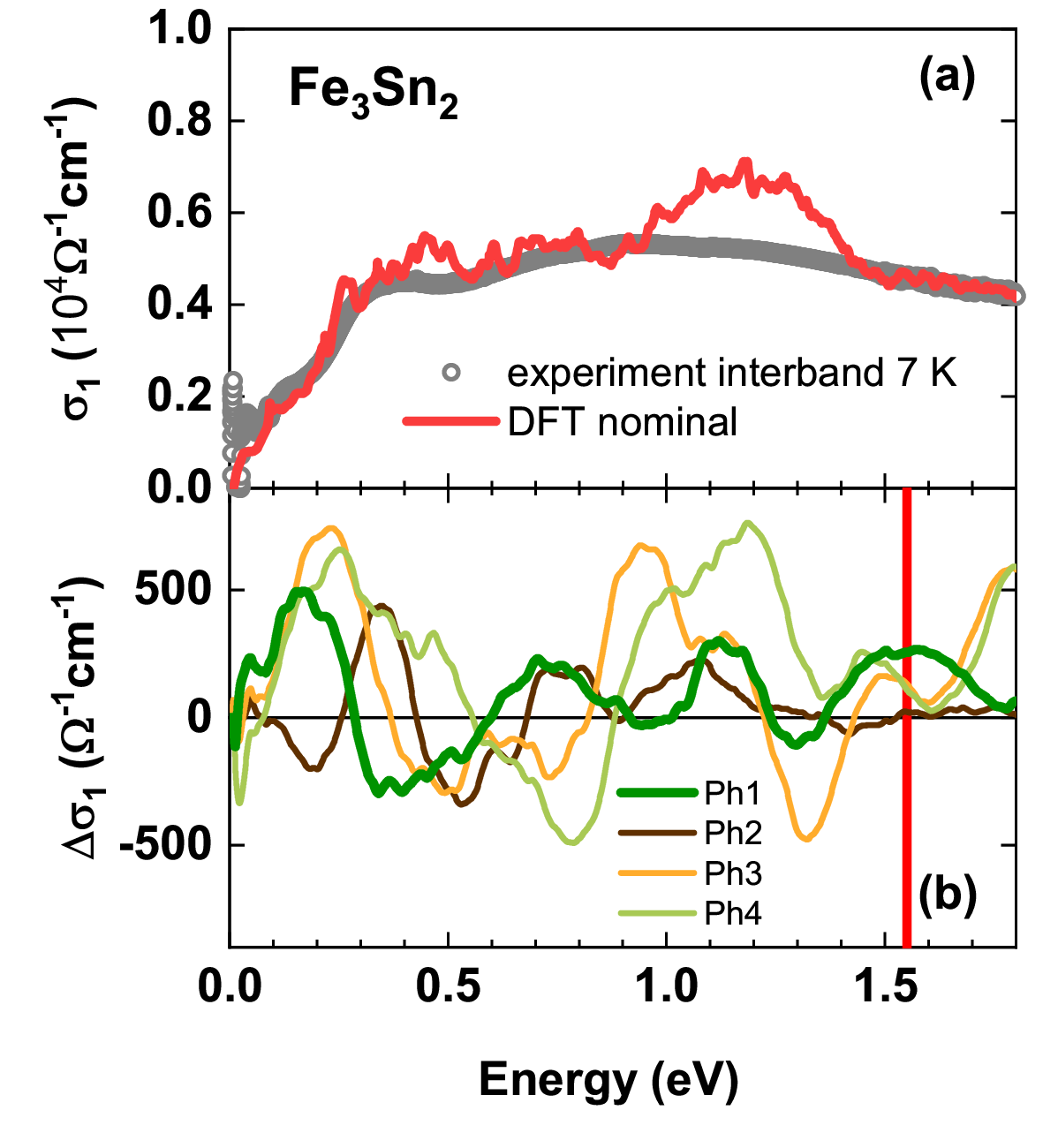}
      \caption{(a) Comparison between experimental and DFT theoretical optical conductivity. (b) Difference in the optical conductivity with respect to the nominal value for all the A$_1g$ modes.}
    \label{DFT_OC}
\end{figure}  

For the DFT calculations, we chose 10~pm as an average displacement. Then, this 10~pm displacement is introduced to the structure given in Table~(\ref{structure}), and the optical conductivity is re-calculated with the new structural parameters. In Fig.~\ref{DFT_OC}(a), the optical conductivity calculated with the non distorted structure is shown as a comparison to the experimental optical conductivity \hyperlink{Biswas2020}{\color{blue}[2]}. The good match with the experiment is obtained only after re-scaling the energy scale of the calculations by /1.4, which is commonly observed for strongly correlated systems. Here the intraband contributions were subtracted from the experimental optical conductivity for a direct comparison with the DFT results. In Fig.~\ref{DFT_OC}(b), the difference in the optical conductivity with respect to the nominal optical conductivity is given for all the $A_{1g}$ modes. As one can immediately notice, at the 800~nm energy range of our pump-probe study the appreciable change in the conductivity only occurs for the Ph1, whereas the changes with other modes are negligible small. This is one of the reasons that we only observe Ph1 in this energy range.

\makeatletter
\renewcommand\@bibitem[1]{\item\if@filesw \immediate\write\@auxout
    {\string\bibcite{#1}{S\the\value{\@listctr}}}\fi\ignorespaces}
\def\@biblabel#1{[S#1]}
\makeatother

\section*{Supplementary references}
\begin{small}

\setlength\parindent{0pt}\hypertarget{Wang2016}[{S1}] Q. Wang, S. Sun, X. Zhang, F. Pang, and H. Lei, Anomalous Hall effect in a ferromagnetic ${\mathrm{Fe}}_{3}{\mathrm{Sn}}_{2}$ single crystal with a geometrically frustrated Fe bilayer kagome lattice, \href{https://doi.org/10.1103/PhysRevB.94.075135} {Phys. Rev. B \textbf{94}, 075135 (2016)}. \\

\setlength\parindent{0pt}\hypertarget{Biswas2020}[{S2}] A. Biswas, O. Iakutkina, Q. Wang, H. Lei, M. Dressel,and E. Uykur, Spin-Reorientation-Induced Band Gap in ${\mathrm{Fe}}_{3}{\mathrm{Sn}}_{2}$: Optical Signatures of Weyl Nodes, \href{https://doi.org/10.1103/PhysRevLett.125.076403} {Phys. Rev. Lett. \textbf{125}, 076403 (2020)}. \\

\setlength\parindent{0pt}\hypertarget{Misochko2004}[{S3}] O. Misochko, M. Hase, K. Ishioka, and M. Kitajima, Observation of an Amplitude Collapse and Revival of Chirped Coherent Phonons in Bismuth, \href{https://doi.org/10.1103/PhysRevLett.92.197401} {Phys. Rev. Lett. \textbf{92}, 197401 (2004)}. \\

\setlength\parindent{0pt}\hypertarget{Zeiger1992}[{S4}] H. Zeiger, J. Vidal, T. Cheng, E. Ippen, G. Dresselhaus, and M. Dresselhaus, Theory for displacive excitation of coherent phonons, \href{https://doi.org/10.1103/PhysRevB.45.768} {Phys. Rev. B \textbf{45}, 768 (1992)}.\\

\setlength\parindent{0pt}\hypertarget{Ultrafast_metals2}[{S5}] P. Allen, Theory of thermal relaxation of electrons in metals, \href{https://doi.org/10.1103/PhysRevLett.59.1460} {Phys. Rev. Lett. \textbf{59}, 1460 (1987)}.\\

\setlength\parindent{0pt}\hypertarget{Ultrafast_metals}[{S6}] R. Schoenlein, W. Lin, J. Fujimoto, and G. Eesley, Femtosecond studies of nonequilibrium electronic processes in metals, \href{https://doi.org/10.1103/PhysRevLett.58.1680} {Phys. Rev. Lett. \textbf{58}, 1680 (1987)}.\\

\setlength\parindent{0pt}\hypertarget{Groeneveld1995}[{S7}] H. Groeneveld, R. Sprik, and A. Lagendijk, Femtosecond spectroscopy of electron-electron and electron-phonon energy relaxation in Ag and Au, \href{https://doi.org/10.1103/PhysRevB.51.11433} {Phys. Rev. B \textbf{51}, 11433 (1995)}.\\

\setlength\parindent{0pt}\hypertarget{wien2k}[{S8}] P. Blaha, K. Schwarz, G.K.H. Madsen, D. Kvasnicka, J. Luitz, R. Laskowski, F. Tran, and L.D. Marks, WIEN2k, An Augmented Plane Wave + Local Orbitals Program for Calculating Crystal Properties (Karlheinz Schwarz, Techn. Universit\"at Wien, Austria), 2018. ISBN 3-9501031-1-2.\\

\setlength\parindent{0pt}\hypertarget{Blaha2020}[{S9}] P. Blaha, K. Schwarz, F. Tran, R. Laskowski, G. K. H. Madsen, and L. D. Marks, WIEN2k: An APW+lo program for calculating the properties of solids, \href{https://doi.org/10.1063/1.5143061} {J. Chem. Phys. \textbf{152}, 074101 (2020)}.\\

\setlength\parindent{0pt}\hypertarget{pbe96}[{S10}] J. Perdew, K. Burke, and M. Ernzerhof, Generalized Gradient Approximation Made Simple, \href{https://doi.org/10.1103/PhysRevLett.77.3865} {Phys. Rev. Lett. \textbf{77}, 3865 (1996)}.\\

\setlength\parindent{0pt}\hypertarget{optic}[{S11}] C. Ambrosch-Draxl and J. Sofo, Linear optical properties of solids within the full-potential linearized augmented planewave method, \href{https://doi.org/10.1016/j.cpc.2006.03.005} {Computer Physics Communications \textbf{175}, 1 (2006)}.\\

\setlength\parindent{0pt}\hypertarget{Stevens2002}[{S12}] T. Stevens, J. Kuhl, and R. Merlin, Coherent phonon generation and the two stimulated Raman tensors, \href{https://doi.org/10.1103/PhysRevB.65.144304} {Phys. Rev. B \textbf{65}, 144304 (2002)}.\\

\end{small}

\end{document}